# A new promising material for biological applications: multi-level physical modification of AgNPs-decorated PEEK

Jana Pryjmaková [1], Daniel Grossberger [1], Anna Kutová [1], Barbora Vokatá [2], Miroslav Šlouf [3], Petr Slepička [1,*], and Jakub Siegel [1]

[1] Department of Solid-State Engineering, University of Chemistry and Technology Prague, 166 28 Prague, Czech Republic; jana.pryjmakova@vscht.cz (J.P.), dan.grossberger@seznam.cz (D.G.), kutovaa@vscht.cz (A.K.), jakub.siegel@vscht.cz (J.S.)

[2] Department of Microbiology, University of Chemistry and Technology Prague, 166 28 Prague, Czech Republic; vokataa@vscht.cz

[3] Institute of Macromolecular Chemistry, Academy of Sciences of the Czech Republic, Heyrovského nám. 2, 162 06 Prague, Czech Republic; slouf@imc.cas.cz

* Correspondence: petr.slepicka@vscht.cz; Tel.: +420-220-445-162

**Abstract: :** In the case of polymer medical devices, the surface design plays a crucial role in contact with human tissue. The use of AgNPs as antibacterial agents is well known; however, their anchoring into the polymer surface can still be investigated. This work describes the change in surface morphology and behaviour in the biological environment of polyetheretherketone (PEEK) with immobilised AgNPs after different surface modifications. The initial composites were prepared by immobilisation of silver nanoparticles from a colloid solution into the upper surface layers of polyetheretherketone (PEEK). The prepared samples (Ag/PEEK) had a planar morphology and were further modified with a KrF laser, a GaN laser, and Ar plasma. The samples were studied using the AFM method to visualise changes in surface morphology and to obtain information on the height of the structures and other surface parameters. Comparative analysis of the nanoparticles and polymers was performed using FEG-SEM. The chemical composition of the surface of the samples and optical activity were studied by XPS and UV-Vis spectroscopy. Finally, drop plate antibacterial and cytotoxicity tests were performed to determine the role of Ag nanoparticles after modification and suitability of the surface, which are important for the use of the resulting composite in biomedical applications.

## 1. Introduction

In recent years, the field of biomedical engineering has witnessed remarkable advances in material science, particularly in the development of novel surface modification techniques to enhance the functionality and performance of biomedical devices. One such material that has gained significant attention is polyetheretherketone (PEEK). With its excellent mechanical properties, biocompatibility, and resistance to harsh environments, PEEK has emerged as a promising candidate for various biomedical applications [1-3]. However, to further expand its potential in the healthcare sector, researchers have been exploring strategies to improve its antibacterial properties, as microbial colonisation remains a persistent challenge in medical devices.

The anchoring of AgNPs to the polymer surface can be achieved through various methods, including chemical reduction, physical deposition, or *in situ* preparation [4-7]. One of the effective methods for anchoring AgNPs is their immobilisation into the polymer near the surface with an excimer laser. This kind of composite represents a long-term source of Ag+ ions [8]. Once AgNPs are successfully immobilised, surface modification techniques are employed to further enhance the material's properties. These techniques can include plasma treatment, laser ablation, chemical functionalisation, or a combination of these approaches.

Plasma treatment is a widely used surface modification technique that involves subjecting the PEEK surface to a low-temperature plasma. This process modifies the chemistry and topography of the surface, resulting in improved wettability and increased surface energy [9]. Plasma treatment can also introduce functional groups that facilitate the attachment of biomolecules and promote cell interactions, making the modified surface more biocompatible. Another effective surface modification method is laser ablation, which involves the use of high-energy lasers, such as KrF excimer lasers, to remove material from the PEEK surface. Laser ablation creates nanostructures and increases surface area, providing an ideal substrate for AgNPs immobilisation [8, 10]. This technique allows precise control over surface morphology and can be tailored to specific applications, resulting in improved integration with surrounding tissues. In addition, the combination of nanometals and plasma or laser treatment represents a synergistic effect on antibacterial activity [11-16].

The combination of AgNP immobilisation and subsequent surface modification holds great potential for a wide range of biomedical applications. By leveraging the unique properties of AgNPs and the benefits of surface modification techniques, researchers aim to develop PEEK-based materials with superior antibacterial capabilities, reduced risk of infections, and enhanced biocompatibility [17-21]. These advances pave the way for the design and fabrication of biomedical devices, such as implants, catheters, and prosthetics, that offer improved performance and patient outcomes.

In this paper, we explore the recent advances in surface modification techniques following AgNP immobilisation on PEEK. We go into the underlying principles of each technique, their resulting effects on surface morphology, and improvements in antibacterial properties and biocompatibility. Furthermore, we discuss the potential applications of these modified surfaces in various biomedical fields, highlighting their significance in combating bacterial colonisation and advancing the development of safer and more effective medical devices.

## 2. Materials and Methods

### 2.1. Synthesis of AgNPs

Colloid silver nanoparticles (AgNPs) were prepared by electrochemical synthesis. Two silver electrodes (5.0 × 1.1 × 0.2 $cm^3$, purity 99.95%, Safina a.s., Czech Republic) were immersed perpendicularly to each other with a distance of 5 cm in sodium citrate dihydrate solution (volume 120 ml, concentration 1 $mmol·l^{-1}$, Sigma-Aldrich Co., USA) and powered by a DC power supply (voltage 15 V, PS-305D). Synthesis was carried out for 20 min under vigorous magnetic stirring (750 $r·min^{-1}$, IKA C-MAG HS 7, Sigma-Aldrich Co., USA). Subsequently, 100 ml of colloid solution was decanted and filtered in an Erlenmeyer flask and kept in a dark place for 24 h. The next day, the concentration of AgNPs solution was determined by atomic absorption spectroscopy and diluted to a work concentration of 30 $mg·l^{-1}$.

### 2.2. Preparation of samples

AgNPs were immobilized into polyetheretherketone surface (PEEK, thickness 50 μm, density 1.3 $g·cm^{-3}$, Goodfellow Cambridge Ltd., UK) with polarized monochromatic light (wavelength 248 nm, UV-grade fused silica, model PBSO-248-100) from the KrF excimer laser (COMPex PRO 50 F, Coherent Inc., USA). The parameters were set as follows: 6000 pulses, pulse duration 20-40 ns, frequency 10 Hz, laser fluence 10 $mJ·cm^{-2}$, aperture 5 × 10 $mm^2$. The immobilisation was carried out with preservation of the planar surface.

The prepared samples with immobilised nanoparticles (Ag/PEEK) were irradiated with a KrF excimer laser and GaN laser. A group of samples (Ag/PEEK/KrF7,

Ag/PEEK/KrF13) was irradiated with a KrF excimer laser under the same conditions as described above with laser fluence of 7 and 13 mJ·cm$^{-2}$. The second group (Ag/PEEK/GaN60, Ag/PEEK/GaN240) was irradiated with a GaN laser (wavelength 405 nm, objective 50x, zoom 6x), which was part of the Olympus LEXT OLS3100 confocal laser scanning microscope (Olympus Corporation, Japan). Exposure times were 60 and 240 s. These samples had a limitation for some analyses, since the modified area was 30×50 μm.

The other method for surface modification was plasma treatment, which was performed using a Sputter Coater, model SCD 050, BAL-TEC set in *etching* mode. The conditions of the experiment were: pressure 4-6 Pa, gas argon, discharge power 8 W, electrode distance 50 mm. The samples were exposed to plasma for 60 and 240 s and are denoted as Ag/PEEK/P60, Ag/PEEK/P240 in text.

*2.3. Analytical methods*

To determine the concentration of synthesised nanoparticles, atomic absorption spectrometry (AAS) was used. The determination was made with the Agilent 280FS AA flame atomiser spectrometer (Agilent Technologies, Japan). The measurement error was 4%.

Atomic force microscopy (AFM) was used to study the surface morphology of Ag/PEEK samples after individual modifications. Measurements were made on a Dimension ICON device, Bruker Corp., USA, using the so-called ScanAsyst® *tapping mode in air*. The analysis was carried out with a SANASYST-AIR probe with a silicon tip and a SiN cantilever (Bruker Corp., USA with an elasticity constant of 0.4 N·m$^{-1}$ and a natural frequency of 70 kHz). Data were acquired at a scan rate of 0.5 Hz and evaluated using the NanoScope® analysis programme. This step also obtained information on the mean surface roughness ($R_a$), the periodicity of the structures ($L$), and the value of the surface area difference (SAD) parameter. Periodicity was measured at five different positions and arithmetical means and standard deviation were calculated.

The visualisation of the surface and arrangement of AgNPs was realized with the MAIA 3 field emission gun scanning electron microscope (FEG-SEM), TESCAN, Czech Republic. Before analysis, the samples were fixed with double-sided carbon tape (Cristine Groepl, Austria) on a brass stub and provided with a carbon layer in a JEE-4C vaporiser (JEOL, Akishima, Japan). Visualisation took place in high-resolution mode with a secondary electron detector at an accelerating voltage of the primary electron beam of 3 kV.

Information about the chemical composition of the samples was obtained from X-ray photoelectron spectroscopy (XPS). The samples were analysed in the ESCAProbeP Omicron Nanotechnology spectrometer (Scienta Omicron, Germany) with a monochromatic X-ray source with an energy of 1,486.7 eV and a pressure of 2·10$^{-8}$ Pa at a take-off angle of 90° (perpendicularly). Data were processed as graph.

Because silver nanoparticles were used, the optical activity of the prepared samples was studied by UV-Vis spectroscopy. Absorption was measured at the Lambda 25 spectrometer (Perkin Elmer, USA) in the range of wavelength 350-700 nm with the scanning rate set at 240 nm·min$^{-1}$. The absorption spectra were obtained from the PerkinElmer UV WinLab software.

*2.4 Biological tests*

An antibacterial activity study was carried out on Ag/PEEK, Ag/PEEK/KrF7, and Ag/PEEK/KrF13 using drop plate tests estimated on counting viable bacteria.[22] In this work, two bacterial strains were chosen: the gram-positive bacteria (G+) *Staphylococcus aureus* (*S. aureus*; CCM 4516) and the gram-negative bacteria (G-) *Escherichia coli* (*E. coli*; CCM 4517). Firstly, the inoculum was prepared using one colony of each bacterial strain into 4 ml (*S. aureus*) or 25 ml (*E. coli*) of Luria-Bertani (LB) liquid medium. The inocula were incubated overnight on an orbital shaker at 37 °C. The next day, they were diluted into phosphate buffered saline (PBS) at a final concentration of approximately 12·10$^3$ (*S. aureus*) and 2·10$^3$ (*E. coli*) per ml. The samples were then immersed in 2 ml of bacterial suspension in test tubes. After 3 and 24 hours, five drops (25 μl) from each tube were dripped onto plate count agar (PCA). The culture was carried out over night at 37 °C. The next day, colony formation units (CFUs) were counted and the arithmetical means and standard deviations were calculated. All experiments were carried out under sterile conditions.

For cytotoxicity tests, the Ag/PEEK, Ag/PEEK/KrF7, and Ag/PEEK/KrF13 were chosen, as samples with very different surface morphology. The cytotoxicity effect was observed in primary lung fibroblasts (MRC-5, American Tissue Culture Collection, Manassas, VA, USA) using a resazurin assay [23]. The experimental setup was taken over from our study dealing with Au nanowires [24].

## 3. Results

### 3.1. Surface Characterisation

The study of surface morphology after laser modification with KrF laser was performed on Ag/PEEK/KrF7, Ag/PEEK/KrF13; with GaN laser Ag/PEEK/GaN60 and Ag/PEEK/GaN240. Scans with visible changes in surface morphology are shown in Figure 1 and adequate surface parameters such as $R_a$, $L$, and *SAD* are summarised in Table 1. In Figure 1 (Ag/PEEK) it is obvious that the immobilisation of AgNPs in the near-surface layer of PEEK was successful and the planar surface was preserved as expected [10]. Subsequent modification with the KrF laser dramatically changed the surface morphology. On the surface of Ag/PEEK/KrF7 and Ag/PEEK/KrF13, nanostructures occurred. Nanostructures formed on the polymer substrate are known as LIPSS (Laser Induced Polymer Surface Structures), which can be arranged in ripples or in globular structures [24, 25]. These structures arise after laser irradiation of polymer substrates, which can absorb UV radiation due to their structures [26].

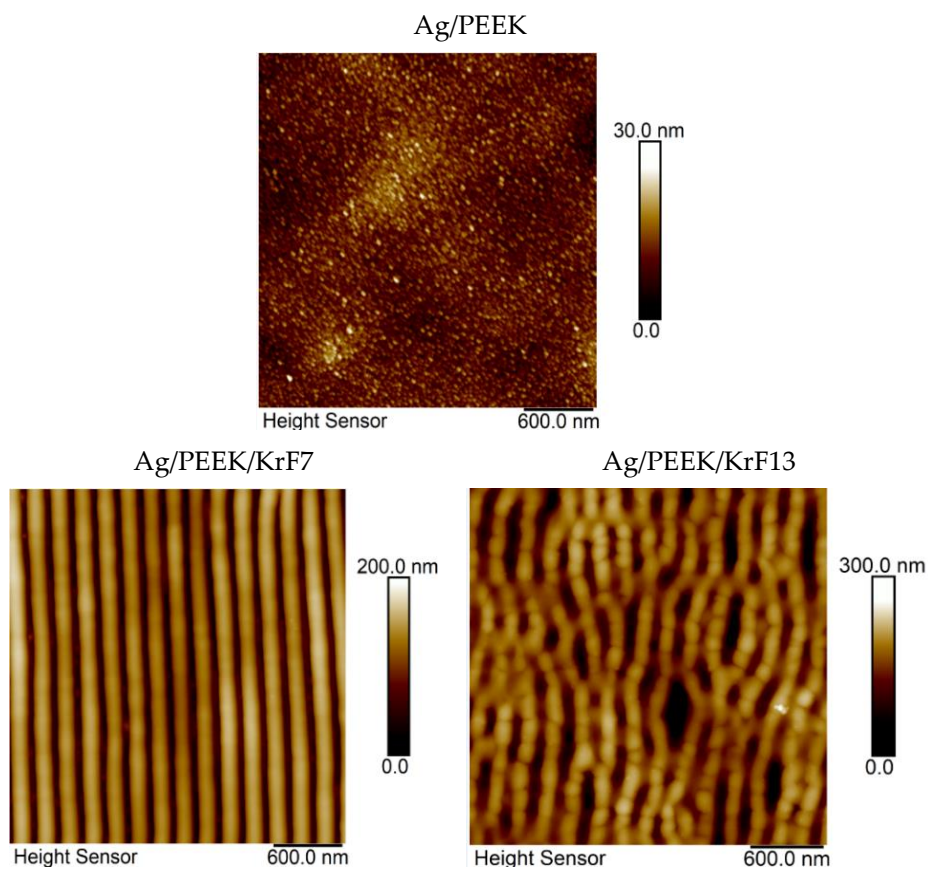

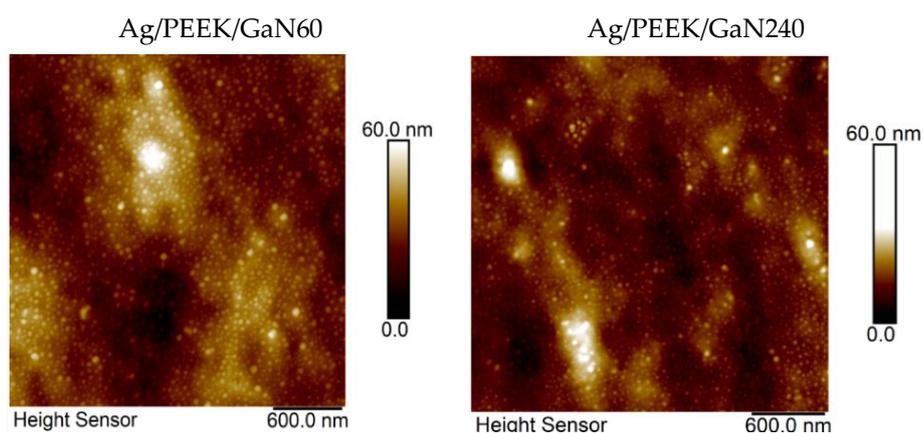

**Figure 1** Images of the Ag/PEEK surface after KrF laser irradiation with 7 (Ag/PEEK/KrF7) and 13 mJ·cm$^{-2}$ (Ag/PEEK/KrF13) and after GaN laser irradiation for 60 (Ag/PEEK/GaN60) and 240 s (Ag/PEEK/GaN240) obtained from AFM.

The laser fluence of 7 mJ·cm$^{-2}$ caused the periodically organized ripples with periodicity of 210 nm. In the case of Ag/PEEK/KrF13, the ripples started degradation and transformed into globular structures, which are typical for higher laser fluence [27]. As LIPSS formed, the $R_a$ reached about 30 nm and the SAD increased significantly. However, AgNPs were hidden under the nanostructures.

**Table 1** Surface parameters such as mean surface roughness ($R_a$), surface area difference (SAD), and periodicity (*L*) of selected samples.

| Sample | $R_a$ (nm) | SAD (%) | *L* (nm) |
|---|---|---|---|
| PEEK | 2.4 | - | - |
| Ag/PEEK | 3.6 | 4.6 | - |
| Ag/PEEK/KrF7 | 31.3 | 52.7 | 210±8 |
| Ag/PEEK/KrF13 | 28.6 | 34.9 | 223±4 |
| Ag/PEEK/GaN60 | 6.8 | 2.3 | - |
| Ag/PEEK/GaN240 | 3.1 | 1.1 | - |

Although the KrF excimer laser with a wavelength of 248 nm created nanostructures on the surface of Ag/PEEK, irradiation with the GaN laser with a wavelength of 405 nm led to material transport, however, without self-assembling. After a 60 s irradiation time, the surface was seeded with larger nanoparticles, after 240 s the small nanoparticles became more visible and most of the large ones were gone. This phenomenon can be related to the distribution of AgNPs through the polymer surface depending on their size [10]. The mass transfer of the polymer gradually uncovers each layer of immobilised nanoparticles, which is in correlation with the values of $R_a$ and SAD. Samples modified with Ar plasma could not be studied with AFM because of the interaction between the charged surface and a silicon tip.

To understand the processes that take place on the polymer surface, the samples were visualised with scanning electron microscopy. The images of Ag/PEEK after each modification are shown in Figures 2-4, where white to light grey represents AgNP and darker shades of grey represent the polymer substrate. The control samples of Ag/PEEK had homogeneously distributed AgNPs on the surface, which is in agreement with the results obtained from AFM. After KrF laser irradiation, the polymer nanostructures occurred without nanoparticles. One can see that the nanoparticles are hidden under the structures (Ag/PEEK/KrF7). However, the smallest AgNPs were transferred to the surface of the nanostructures after laser irradiation with 13 mJ·cm$^{-2}$ (Ag/PEEK/KrF13).

In Figure 3, it can be seen that surface modification with the GaN laser exposed nanoparticles immobilised deeper in the polymer surface. The surface of Ag/PEEK/GaN60 sample contents AgNPs with various sizes, while the surface of Ag/PEEK/GaN240 was covered with small nanoparticles. On closer observation, it is possible to notice the dark colouration of the background around the nanoparticles (represent polymer), which means that irradiation penetrated deeper into the composite with increasing time of irradiation. In that case, the content of AgNPs decreased, and the polymer increased. Similarity can be found in the study of Elashnikov et al. [28], which deals with the irradiation of Ag-doped PMMA fibres with a laser of 405 nm wavelength.

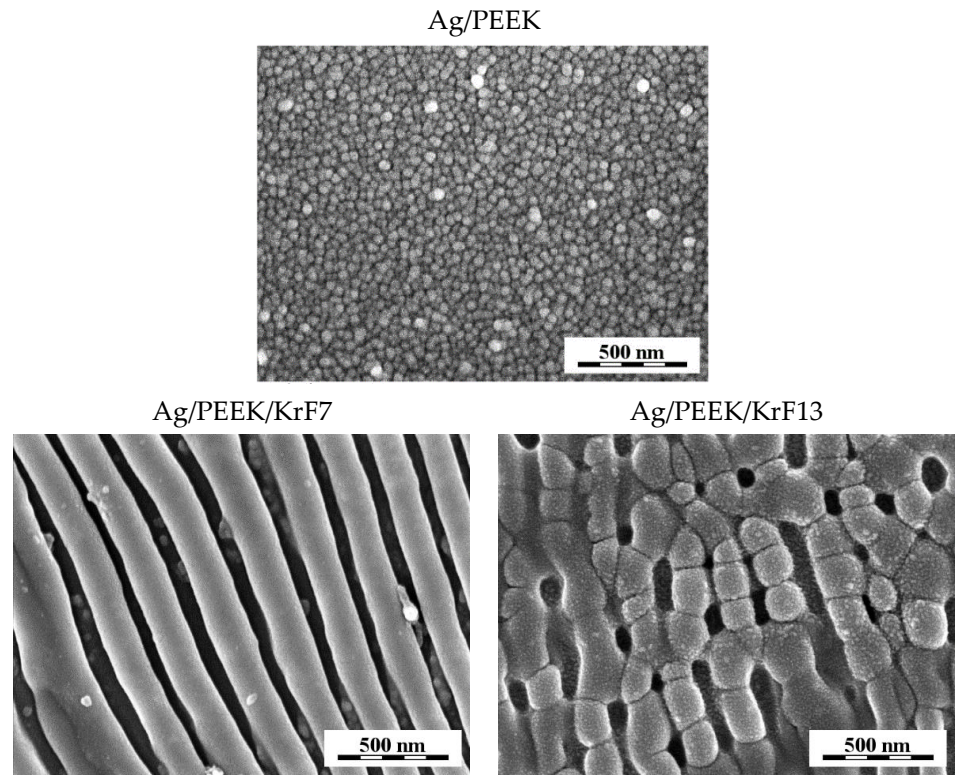

**Figure 2.** Micrographs of Ag/PEEK, Ag/PEEK/KrF7 and Ag/PEEK/KrF13 depicting nanostructures formed after KrF laser irradiation.

As LIPSS did not form in this case, the wrinkling of the material had occurred at the edge of the modified area, which is shown in Figure 3 (Ag/PEEK/GaN60-edge, Ag/PEEK/GaN240-edge). The wrinkling probably was caused by photoactivation of AgNPs, as the wavelength of 405 nm is close to their LPSR [11, 29]. The UV irradiation of AgNPs can heat them to 130 °C, so it has an effect on the transfer of polymer material as the temperature is close to the PEEK glass transition temperature [10, 30]. Nevertheless, ablation of the material may not be excluded.

Here we get to the first results of Ag/PEEK after the Ar plasma modification. Figure 4 contains micrographs of Ag/PEEK/P60 and Ag/PEEK/P240 surfaces. The analysis did not show any significant differences between Ag/PEEK and Ag/PEEK/P60. Interestingly, on Ag/PEEK/P240, visible space between nanoparticles was observed, which could be explained by oxidation or material ablation. In summary, these results show that the PEEK is the most susceptible to the KrF laser modification in terms of surface morphology.

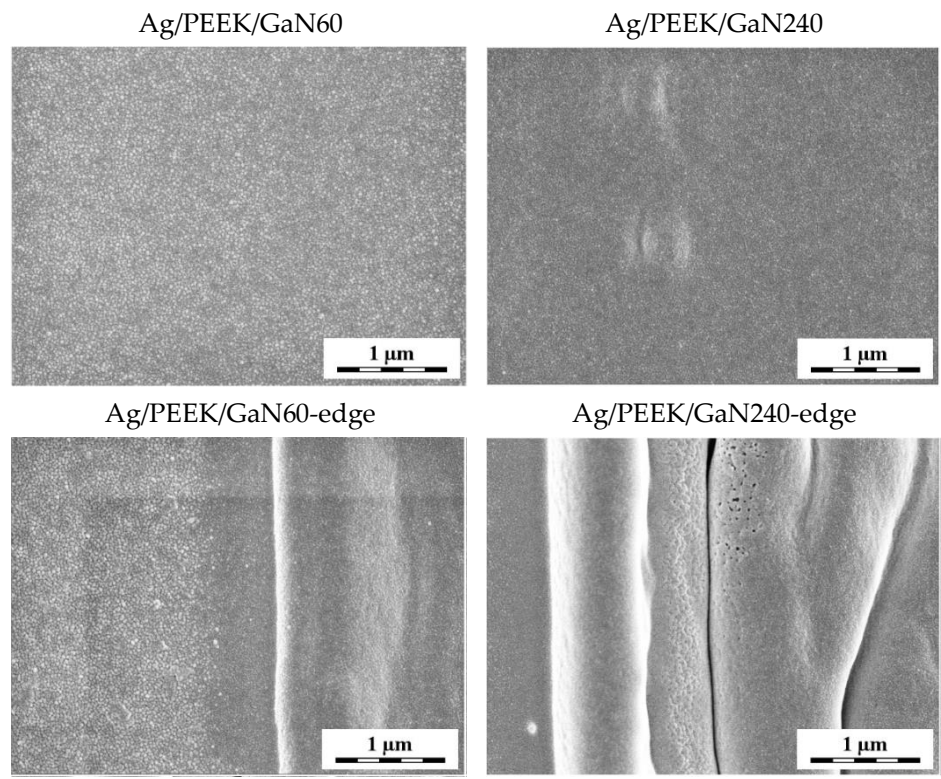

**Figure 3.** Micrographs of Ag/PEEK/GaN60 and Ag/PEEK/GaN240 with visible AgNPs and additional images of the interface unmodified and modified surface of Ag/PEEK noted as Ag/PEEK/GaN60-edge and Ag/PEEK/GaN240-edge

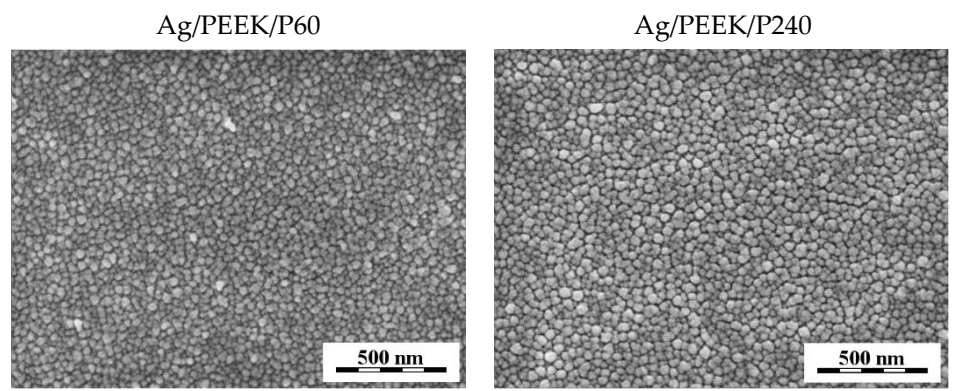

**Figure 4.** Micrographs of Ag/PEEK/P60 and Ag/PEEK/P240 after exposure to Ar plasma

To detect any changes in surface chemistry, the concentrations of elements in the surface of Ag/PEEK, Ag/PEEK/KrF7, Ag/PEEK/KrF13, Ag/PEEK/P60 and Ag/PEEK/P240 were determined, which are summarized in Figure 5. The PEEK was used as a pristine. The XPS confirms the presence of AgNPs in the surface as the take-off angle was 90°, which can obtain information from the 20 nm depth [31]. Surface irradiation with the KrF excimer laser significantly decreased content of the silver from 20% to 6%, in the case of 7 mJ·cm$^{-2}$ it is even lower. The decrease was caused by the creation of LIPSS that completely cover nanoparticles. As we expected from FEG-SEM analysis, nanoparticles occurred in a small amount in the surface of the LIPSS after higher laser fluence. On the other hand, the concentration of oxygen slightly increased. The interaction of laser irradiation and polymer leads to many complex processes; one of them is the partial oxidation of the polymer [32].

The plasma modified surface brings more interesting results. Although the surface morphology of Ag/PEEK and Ag/PEEK/P60, Ag/PEEK/P240, looked pretty similar, the chemistry of these surfaces was different. The concentration of silver was reduced by half of its original value. Plasma treatment led to oxidation of the polymer, which is also evidenced by the increase in oxygen with prolonged treatment [33]. However, it is not possible to say with certainty whether only the polymer or the nanoparticles were oxidised. We expected that nanoparticles would not provide any changes in chemical composition as Ar plasma was used, regardless of the impurities of 4-6% being present and the oxidation of nanoparticles is possible. The chamber was cleaned before each modification; however, impurities such as iron and silicon were still present. The source of iron and silicon could be the sputter coater itself, as most of the components were made of iron alloys and the cylinders were made of silicon glass. Moreover, these residues can be the cause of limitations in AFM measurement.

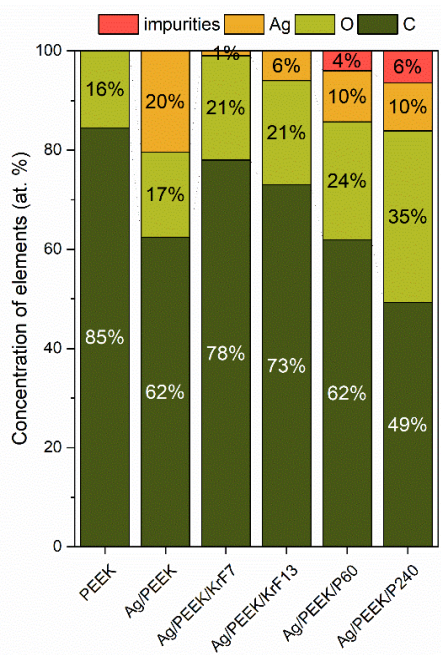

**Figure 5.** Graph showing the concentrations of carbon (C 1s), oxygen (O 1s), and silver (Ag 3d) in the surface of the Ag/PEEK composite and its modified versions.

The UV-Vis spectra of the chosen samples are shown in Figure 6. Since the controls did not show optical activity, all samples contained silver nanoparticles exhibited an optical absorption of about 450 nm. Similar trends have been reported by Zeng et al. in their work on the silver-aromatic polymer composite AgNPs/PS (polystyrene) and AgNPs/AP (acrylonitrile-polystyrene), where the absorption band was 450-460 nm [34]. As can be seen in Figure (6. A), composites irradiated with the KrF laser show a decreasing trend of absorbance. Peaks became narrower and more distinct with increasing laser fluence. The cause of this phenomenon can be explained by the different behaviour of the nanoparticles. In the case of Ag/PEEK, silver nanoparticles appeared as a continuous layer. The layer of silver nanostructures that creates it has a specific absorption spectrum depending on its thickness or conditions of preparation [35, 36]. With increasing laser fluence, nanoparticles became more separate (see Figure 2), so their optical response is close to that of colloid nanoparticles. The results of plasma-treated samples show the opposite trend of optical absorbance. After modification for 60 s, the concentration of silver dramatically decreased, which was reflected in a non-committal peak at 445 nm. Nonetheless, prolonged plasma treatment did not produce any dramatic changes.

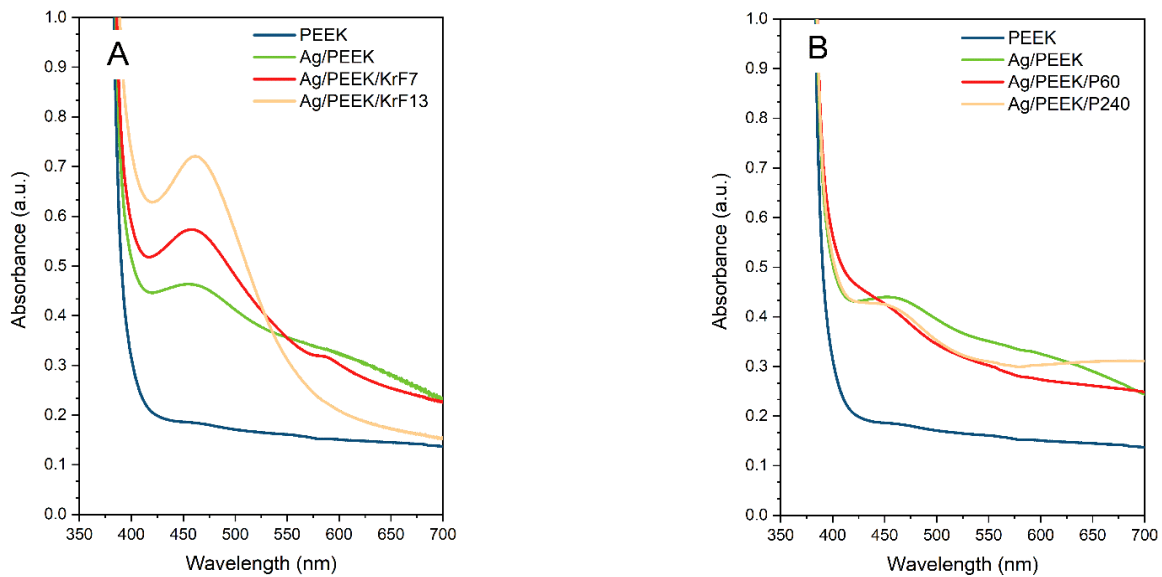

**Figure 6.** UV-Vis spectra of composites after (A) KrF laser irradiation and (B) Ar plasma treatment.

*3.2. Antibacterial tests*

Because we were interested in antibacterial properties of prepared composites, we performed drop plate tests on Ag/PEEK, Ag/PEEK/KrF7, and Ag/PEEK/KrF13 due to their different concentrations of silver. In this work, tests were performed on *Escherichia coli* and *Staphylococcus aureus*, as one of the most common strains of bacteria, which caused catheter-associated urinary tract infections (CAUTI) and are responsible for the most of medical device-related infections [37-39]. The results of antibacterial activity are shown in Figure 7. It is obvious that we achieved the necessary effect. After 3 h of incubation time, the samples did not show antibacterial activity against both strains of bacteria. However, the CFU of *E. coli* decreased slightly after interaction with all samples with silver. The significant decrease was in case of Ag/PEEK/KrF13, which may be caused by the combination of silver nanoparticles and structural surface morphology. Similar trends have been reported by Kaimlová et al. [40] in their work on silver nanowires supported on laser-patterned polyethylennaphthalate. More interesting results occurred after 24 h of incubation time. The antibacterial effect was revealed for all Ag-doped samples against both strains of bacteria. Antibacterial effects on *E. coli* and *S. aureus* were also observed in the study of Deng et al. [20], where PEEK was used as a 3D printed implant with AgNPs. Moreover, total inhibition of the bacteria grown was crucial in the case of *S. aureus*, which is known for its antibiotic resistance [41].

*3.3. Cytotoxicity tests*

As prepared composites are designed for biomedical applications, the cytotoxicity effect of Ag/PEEK, Ag/PEEK/KrF7, and Ag/PEEK/KrF13 was tested on primary lung fibroblasts. The results of these tests are presented in Figure 8. From the graph, it is obvious that Ag/PEEK showed a weak toxicity effect after 24 h of incubation time. A moderate cytotoxic effect was observed after a longer incubation time, which is alarming; therefore, Ag/PEEK is not a suitable candidate for biomedical applications. Cytotoxicity occurred due to the high concentration of silver nanoparticles and the low surface roughness, since it is known that surface-modified PEEK is not toxic to cell adhesion and proliferation [26]. However, this composite can be used as a support material for the packaging of medical devices with poor sterilisation or storage conditions. The most significant observation emerging from these tests was the non-cytotoxic effect of the samples after KrF laser modification after 24 and 48 h of incubation time. Cells adhered and proliferated on the surface of Ag/PEEK/KrF7 and Ag/PEEK/KrF13 due to the diverse surface morphology, because the surface morphology and roughness are important for cell adaptation [42]. Not only surface morphology plays a role in cell adhesion and proliferation, but also form and concentration of silver. In the case of Ag/PEEK, AgNPs are distributed all over the surface of PEEK, so cells practically become in contact with the silver layer, which is toxic to them. After KrF laser modification, the nanostructures occurred and the concentration of silver decreased, which offers more appropriate conditions for cell adhesion. Similar cytotoxicity tests were performed on polyethylennaphthalate (PEN) also with LIPSS decorated with silver nanowires (AgNWs) [43]. Unfortunately, cytotoxic effect tests showed strong cytotoxicity in the case of AgNWs/PEN even after 72 h of incubation time.

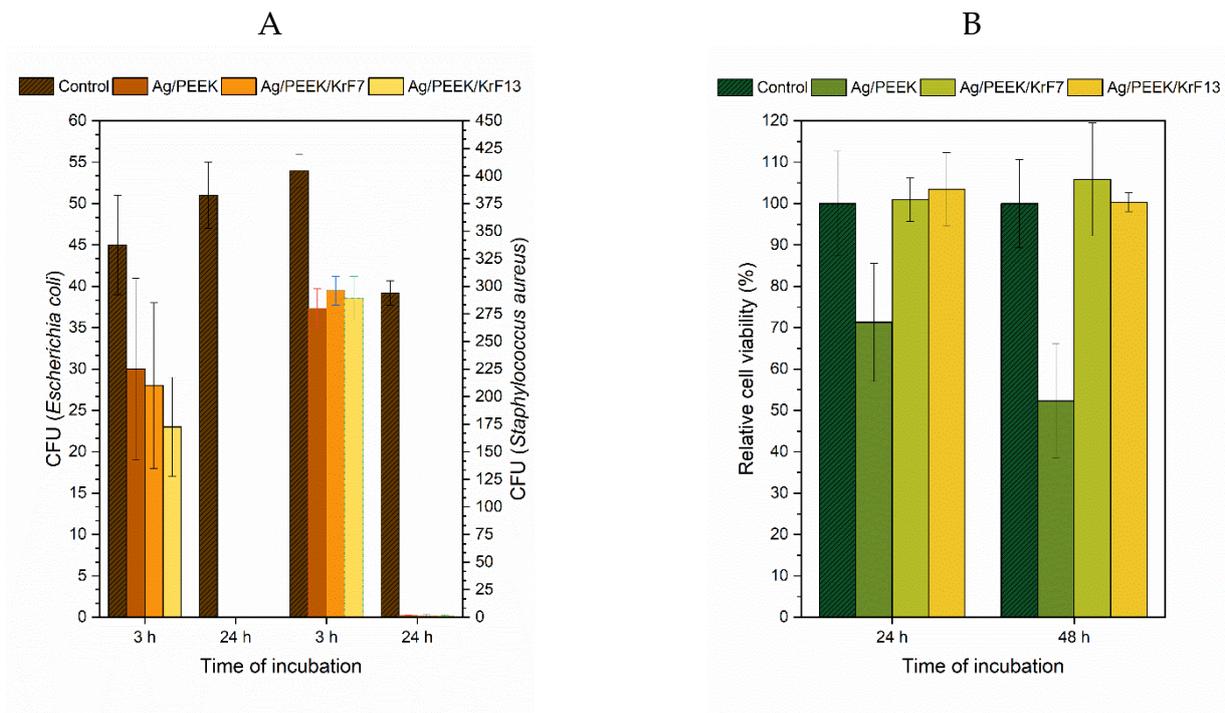

**Figure 7.** Results from biological tests shown (A) data from drop plate tests performed against *Escherichia coli* (left y-axis) and *Staphylococcus aureus* (right y-axis) after 3 and 24 h of incubation, expressed in CFU and (B) non-cytotoxic effect of Ag/PEEK/KrF7 and Ag/PEEK/KrF13 samples on primary lung fibroblasts cultivated for 24 and 48 h, expressed by rel. cell viability (%).

## 4. Conclusions

In summary, the investigation into polyetheretherketone (PEEK) with immobilised silver nanoparticles (AgNPs) has unveiled a promising avenue for advanced material engineering. Examination of LIPSS induced by KrF laser treatment demonstrates the potential for precise surface modifications that can have substantial implications in improving material properties. Moreover, these structures can be explored as a basis for direction control grown from specific cells (myocytes, or bone cells). The mass-transfer effects observed after GaN laser treatment offer a novel avenue for controlled material manipulation. Although plasma treatment resulted in non-significant changes in surface morphology, this finding underlines the nuanced nature of material modifications and the need for further investigation and optimisation. XPS and UV-Vis analysis revealed stability of the surface chemistry and optical activity of KrF laser treated composites in contrast to plasma modified. The discovery of complete inhibition of *Escherichia coli* and *Staphylococcus aureus*, without cytotoxic effects, highlights the promise of these materials in medical and antibacterial applications. Collectively, these findings suggest that these developments in material science hold significant potential for various medical applications, making it an attractive candidate for medical and healthcare applications, where the complete inhibition of *E. coli* and *S. aureus* has significant implications. Further research and exploration in these areas promise to bring about even more exciting possibilities for the future.

**Author Contributions:** Conceptualisation, visualisation and writing – original draft, sample preparation, J.P.; sample preparation, UV-Vis measurement, D.G; sample preparation, antibacterial tests, cytotoxicity tests, A.K. and B.V; FEGSEM analysis, M.Š.; funding acquisition, data curation, P.S.; supervision, methodology, writing-review and editing, AFM analysis, J.S. All authors have read and agreed to the published version of the manuscript.

**Funding:** This work was supported from the grant of Specific university research – grant No. A2_FCHT_2023_003, Czech Science Foundation under the project No. 22-17346S, and Project OP JAK_Mebiosys, No CZ.02.01.01/00/22_008/0004634, of the Ministry of Education, Youth and Sports, which is co-funded by the European Union.

**Data Availability Statement:** The data presented in this study are available at https://doi.org/10.5281/zenodo.10234756.

**Acknowledgments:** J.P. is thankful to Filip Šturc for his personal support during the experimental work.

**Conflicts of Interest:** The authors declare no conflict of interest.